%% file: cikm2016.tex
\documentclass{sig-alternate-05-2015}

\usepackage{times}
\usepackage{latexsym}
\usepackage{amsmath}
\usepackage{amssymb}
\usepackage{multirow}
\usepackage{url}
\usepackage{booktabs}
\usepackage{pgfplots}
\usepackage{color}
\usepackage{siunitx}
\usepackage{subfloat,subfig}
\usepackage{microtype}
\usepackage{textcomp }
\usetikzlibrary{plotmarks}

\newcommand{\knn}{\mbox{$k$-NN} }
\newcommand{\ttt}{\texttt}

\newcommand{\yahoons}{Yahoo!\hspace{-0.1em}}
\newcommand{\yahoo}{\yahoons{} }
\newcommand{\comprns}{\emph{Comprehensive}}
\newcommand{\compr}{\comprns{} }
\newcommand{\manrns}{\emph{Manner}}
\newcommand{\manr}{\manrns{} }
\newcommand{\stackons}{\emph{Stack Overflow}}
\newcommand{\stacko}{\stackons{} }
\newcommand{\tblfntsz}{\scriptsize}

\newcommand{\modelonens}{Model~1}
\newcommand{\modelone}{\modelonens{} }
\newcommand{\bmmodelonens}{\emph{BM25+Model~1}}
\newcommand{\bmmodelone}{\bmmodelonens{} }
\newcommand{\bmtfns}{\emph{BM25}}
\newcommand{\bmtf}{{\bmtfns{} }}
\newcommand{\tfidfns}{{TF$\times$IDF}}
\newcommand{\tfidf}{\tfidfns{} }
\newcommand{\tfidfcosinens}{\emph{Cosine \tfidfns}}
\newcommand{\tfidfcosine}{\tfidfcosinens{} }
\newcommand{\embedcosinens}{\emph{Cosine Embed}}
\newcommand{\embedcosine}{\embedcosinens{} }
\newcommand{\ms}[1]{\SI{#1}{\milli\second}}
\newcommand{\second}[1]{\SI{#1}{\second}}
\usepackage{hyperref}

\newcommand{\anonymizeurl}[1]{\url{#1}}

\usepackage{todonotes}

\newif\ifProduction
\Productiontrue

\ifProduction
\newcommand{\leoinline}[1]{}
\newcommand{\davidinline}[1]{}
\else
\newcommand{\leoinline}[1]{{\color{red}\textbf{#1}}}
\newcommand{\davidinline}[1]{{\color{green}\textbf{#1}}}
\fi

\begin{document}
\CopyrightYear{2016}
\setcopyright{acmlicensed}
\conferenceinfo{CIKM'16 ,}{October 24-November 28, 2016, Indianapolis, IN, USA}
\isbn{978-1-4503-4073-1/16/10}\acmPrice{\$15.00}
\doi{http://dx.doi.org/10.1145/2983323.2983815}
% --- End of Author Metadata ---

\clubpenalty=10000 
\widowpenalty = 10000

\title{Off the Beaten Path: Let's Replace Term-Based Retrieval with k-NN  Search}

\numberofauthors{4} %  in this sample file, there are a *total*
\author{
\alignauthor
       Leonid Boytsov\\
       \affaddr{Carnegie Mellon University}\\
       \affaddr{Pittsburgh, PA, USA}\\
       \email{srchvrs@cs.cmu.edu}
\alignauthor
       David Novak\\
       \affaddr{Masaryk University}\\
       \affaddr{Brno, Czech Republic}\\
       \email{david.novak@fi.muni.cz}
\alignauthor
       Yury Malkov\\
       \affaddr{\mbox{Institute of Applied Physics RAS}}\\
       %\affaddr{46 Ul'yanov Street}\\
       \affaddr{\mbox{Nizhny Novgorod, Russia}}\\
       \email{yurymalkov@mail.ru}
\and
\alignauthor
       Eric Nyberg\\
       \affaddr{Carnegie Mellon University}\\
       \affaddr{Pittsburgh, PA, USA}\\
       \email{ehn@cs.cmu.edu}
}
% Just remember to make sure that the TOTAL number of authors
% is the number that will appear on the first page PLUS the
% number that will appear in the \additionalauthors section.

\maketitle
\begin{abstract}
Retrieval pipelines commonly rely on a term-based search to obtain candidate records, which are subsequently re-ranked. Some candidates are missed by this approach, e.g., due to a vocabulary mismatch. We address this issue by replacing the term-based search with a generic \knn retrieval algorithm, where a similarity function can take into account subtle term associations. While an exact brute-force \knn search using this similarity function is slow, we demonstrate that an approximate algorithm can be nearly two orders of magnitude faster at the expense of only a small loss in accuracy. A retrieval pipeline using an approximate \knn search can be more effective and efficient than the term-based pipeline. This opens up new possibilities for designing effective retrieval pipelines.
Our software (including data-generating code) and derivative data based on
the \stacko collection is available online.\footnote{\anonymizeurl{https://github.com/oaqa/knn4qa}} 
This revision is a slightly extended version of the respective CIKM'16 paper.
\end{abstract}



%
%  Use this command to print the description
%
\printccsdesc

% We no longer use \terms command


\section{Introduction}
\input intro.tex

\section{Approach}
\label{sec:approach}
\input approach.tex

\section{Main Experiments}
\label{sec:exper}
\input main_exper.tex

\section{Discussion and Related Work}
\label{sec:discuss}
\input discussion.tex

\section{Conclusion}
\label{sec:conclusion}
\input conclusion.tex

\section*{Acknowledgements}
Leonid Boytsov is supported by the Open Advancement of Question Answering Systems (OAQA) group.\footnote{\url{https://oaqa.github.io/}};
David Novak is supported by the Czech Research Foundation project P103/12/G084;
Yury Malkov is supported by the Russian Foundation for Basic Research (project No. 16-31-60104 mol\_a\_dk).

We also thank Di Wang for helping with a Lucene baseline;
Chris Dyer for a discussion of IBM Model~1 efficiency;
Yoav Goldberg, Manaal Faruqui, Chenyan Xiong, Ruey-Cheng Chen for discussions related to word embeddings;
Michael Denkowski for a discussion on approximating alignment scores.

%\newpage
%\footnotesize
%\small
%\bibliographystyle{abbrv}
%\bibliography{cikm2016}

\end{document}

%% file: intro.tex
Due to advances in computing, a full-text search has become a ubiquitous information technology.
However, this technology still largely relies on memorization
of document terms and 
matching them with the query terms provided by a user.

The full-text search is powered by a \emph{term-based} inverted index: 
a classic data structure that links document terms---and sometimes phrases---with their locations in a text collection.
This way of organizing text data traces back to paper book indices
containing alphabetical lists of principal words.
In particular, it was used in a 13th century Bible concordance, long before the computer era \cite{fenlon1913}. 

Modern retrieval systems answer queries in a pipeline fashion. 
First, an term-based inverted index is used to generate a list of candidate documents containing some or all query terms.
Second, this list is refined and re-ranked.
A few highly-ranked documents are then presented to the user.

Re-ranking may be carried out in several steps, where earlier steps employ
cheap ranking functions---such as BM25 \cite{Robertson2004} or language models \cite{ponte1998language}---relying solely on term occurrence statistics.
A final, aggregation, step typically combines numerous relevance signals generated by upstream components.
The aggregation step is often carried out using statistical \emph{learning-to-rank} algorithms \cite{Liu2009}.

This filter-and-refine approach hinges on the assumption that a term-based search
generates a reasonably complete list of candidate documents.
However, this assumption is not fully accurate, in particular, because of a \emph{vocabulary gap}, 
i.e., a mismatch between query and document terms denoting same concepts.
The vocabulary gap is a well-known phenomenon.
Furnas~et~al.~\shortcite{furnas1987vocabulary} showed that, given a random concept,
there is less than a 20\% chance that two randomly selected humans denote this concept using the same term.
Zhao and Callan \shortcite{zhao2010term} found that a term mismatch ratio---i.e., a rate at which a query
term fails to appear in a relevant document---is roughly 50\%.

Furthermore, according to Furnas et al.~\shortcite{furnas1987vocabulary}, 
focusing only on a few synonyms is not sufficient to effectively bridge the vocabulary gap. 
Specifically, it was discovered that, after soliciting 15 synonyms describing a single concept from a panel of subject experts,
there was still a 20\% chance that a new person coined a previously unseen term.
To cope with this problem, Furnas~et al. \shortcite{furnas1987vocabulary} proposed a system of \emph{unlimited term aliases}, 
where potential synonyms would be interactively explored and presented to the user in a \emph{dialog} mode. 
%The order of exploration would depend on a probability that synonyms are associated with original query terms. 

An established \emph{automatic} technique aiming to reduce the vocabulary gap is a \emph{query expansion}. 
It consists in expanding (and sometimes rewriting) a source query using related terms and/or phrases.
For efficiency reasons, traditional query expansion techniques are limited to dozens of expansion terms \cite{Carpineto2012}.
Using hundreds or thousands of expansion terms seems to be infeasible within a framework of the term-based inverted index.
In contrast, we demonstrate that a system of unlimited term aliases can be successfully implemented within a more generic framework of a \emph{$k$-nearest neighbor search} (\knn search).

It has been long recognized that the \knn search shows a promise to make retrieval a \emph{conceptually} simple optimization procedure \cite{Konopnicki2005}. 
This approach may permit a separation of labor between data scientists, focusing on methods' accuracy, and software engineers, focusing on development of more efficient and/or scalable search approaches.
However, the \knn search proved to be a challenging problem due to the curse of dimensionality.
There is empirical and theoretical evidence that this problem cannot be solved both exactly and efficiently in a high-dimensional setting \cite{weber1998quantitative,beyer1999nearest,chavez2001searching,Pestov2012}.
For some data sets, e.g., in the case of vectors with randomly generated elements, 
exact methods degenerate to a brute force search for just a dozen of dimensions \cite{weber1998quantitative,beyer1999nearest}.
Some data sets only ``look'' high-dimensional, but possess properties of low-dimensional data sets, i.e.,
they have a low \emph{intrinsic} dimensionality \cite{korn2001dimensionality,beyer1999nearest,chavez2001searching}.
Unfortunately, textual data seems to be intrinsically high-dimensional. 
For example, 
using the definition of Ch{\'a}vez et al.~\cite{chavez2001searching}, we estimate 
that the intrinsic dimensionality of Wikipedia \tfidf vectors is about 2500 in the case of the metric angular distance.

The curse of dimensionality can be partially lifted by using Locality Sensitive Hashing (LSH) techniques 
\cite{broder1997resemblance,indyk1998approximate,Kushilevitz_et_al:1998}.
There are numerous modifications of LSH, which differ primarily in how they construct families of locality-sensitive functions \cite{wang2014hashing}.
Most of the research focuses on hash functions for well-studied similarities,
such as the Euclidean distance and the cosine similarity.
%For the purpose of the \knn search the angular distance is equivalent to the cosine simlarity,
%which is a popular similarity function for comparing textual data in IR and NLP \cite{petrovic2010streaming,Ture2011,Li2014,Petrovic2012,Moran2016}.\footnote{There is a monotonic mapping from the cosine similarity to the angular distance and vice versa.  This transformation preserves the relative order (but not distances) among nearest neighbors.}
%There are tasks (e.g., a streaming first story detection \cite{allan2000detections}) where the cosine similarity has been reported to be superior to other distance functions.
%However, there is also plenty of evidence that BM25-based similarity functions are substantially more accurate 
%than the cosine-similarity (see, e.g., a paper by
%Whissell and Clarke and references therein \cite{Whissell2013}).
%Specifically, this is true for our data (see \S~\ref{sec:exper}). 

%Furthermore, we find that a linear combination of BM25 and IBM \modelone scores, can be up to 25\% more effective than BM25 alone.
%However, the resulting \emph{expensive to compute} similarity function, henceforth denoted as \bmmodelonens, is \emph{non-symmetric} and \emph{non-metric}. 
%Due to lack of symmetry, one cannot use the non-metric extensions of LSH based on the kernel trick \cite{kulis2009kernelized,Mu2010NonMetricLH}.

In this paper, however, we explore an effective similarity function \bmmodelonens, which is neither metric nor symmetric (see \S~\ref{sec:bm25modelone}).
We demonstrate that it is possible to carry out an efficient and effective \knn search (for \bmmodelonens) using pivoting techniques.
In that, the approximate \knn search is nearly two orders of magnitude faster than the respective exact brute force search.
The \knn search can be 1.5$\times$ faster than Lucene, while being more effective due to bridging the vocabulary gap.

%\leoinline{This needs to be deleted, if Leo can't deliver on this point:}
%We also provide evidence that for our data set it is hard to obtain a similar result using a classic query expansion.

To ease reproducibility,
we make our software (including data-generating code) and derivative data based on the Stack Overflow collection 
available online.\footnote{\anonymizeurl{https://github.com/oaqa/knn4qa}}

%% file: approach.tex
We focus on a task of searching a large collection of answers extracted from a community QA website.
The questions and answers are submitted by real people, who also select \emph{best} answers.
A question and the respective best answer represent one QA pair.
While community QA is an important task on its own, %which is also quite challenging \cite{Fried2015}, 
it is used here primarily as a testbed to demonstrate the potential of the \knn search as a substitute for term-based retrieval.
Due to the curse of dimensionality, we have to resort to approximate searching. 
Note that we need a similarity function that outstrips the baseline method BM25
by a good margin. Otherwise, gains achieved by employing a more sophisticated similarity would be invalidated
by the inaccuracy of the search procedure.

One effective way to build such a similarity function is to learn a generative question-answer \emph{translation} model,
e.g., IBM \modelone \cite{brown1993mathematics}.
However, ``\ldots the goal of question-answer translation is to learn associations between question terms and synonymous answer terms, rather than the translation of questions into fluent answers.'' \shortcite{RiezlerEtAl2007}
The idea of using a translation model in retrieval applications was proposed by Berger et al.~\cite{Berger2000}.
%and is now widely adopted by the IR and QA communities (see, e.g., Surdeanu et al.~\cite{surdeanu2011learning}
%and references therein). 
It is now widely adopted by the IR and QA communities \cite{Echihabi2003,Soricut2006,RiezlerEtAl2007,Xue2008,surdeanu2011learning,Fried2015}.
Linearly combining BM25 and \emph{logarithms} of IBM Model 1 scores %using a simple learning-to-rank algorithm \cite{Metzler2007}
produces a similarity function that is considerably more accurate than BM25 alone (by up to 30\% on our data, see Table~\ref{tab:main}).

Learning IBM \modelone requires a large monolingual parallel corpus. 
In that, the community QA data sets seem to be the best publicly available source of such corpora.
Note that a monolingual corpus can be built from search engine click-through logs \cite{riezler2010query}.
Yet, such data is not readily available for a broad scientific community.
Another advantage of community QA data sets is that they permit a large scale automatic evaluation with sizeable training and testing subsets.

Specifically, we extract QA pairs from the following collections:  
\begin{itemize}
\item L6 - \yahoo Answers \compr version 1.0 (about 4.4M questions);
\item L5 - \yahoo Answers \manr version 2.0 (about 142K questions),
which is a \emph{subset} of L6 created by Surdeanu~al.~\cite{surdeanu2011learning};  
\item \stacko (about 8.8M answered questions).
\end{itemize}
\yahoo Answers collections are available through \yahoo WebScope and can be requested by researchers from accredited universities.\footnote{\url{https://webscope.sandbox.yahoo.com}} For each question, there is always an answer (and the best answer is always present).
The \stacko collection is freely available for download.\footnote{We use  a dump from \url{https://archive.org/download/stackexchange} dated March 10th 2016.} While there are 8.8M answered questions, 
the best answer is not always selected by an asker. Such questions are discarded leaving us with 6.2M questions.

{

Each question has a (relatively) short summary of content, which is usually accompanied by a longer description.
The question summary concatenated with the description is used as a query with the objective of retrieving the corresponding best answer.  
The best answer is considered to be the \emph{only} relevant document for the query.

The accuracy of a retrieval system is measured using standard IR metrics: 
a Mean Reciprocal Rank (MRR), 
a precision at rank one (P@1) and an answer recall measured for the set of top-$N$ ranked documents.
P@1---our main evaluation metric---is equal to a fraction of queries where the highest ranked document is a true best answer to the question.

We process collections by removing punctuation, extracting tokens and term lemmas
using Stanford CoreNLP \cite{manning2014} (instead, one can use any reasonably accurate tokenizer and lemmatizer).  All terms and lemmas are lowercased; stopwords are removed. 
Note that we keep \emph{both} lemmas and original terms.
In \stacko we remove all the code (the content marked by the tag \ttt{code}).

Each collection is randomly split into several subsets, which
include training, two development (\emph{dev1} and  \emph{dev2}), and testing subsets.
%For testing purposes though we use only the first 10K questions.
In the case of \compr and \stackons, there is an additional subset that is used to learn IBM \modelonens.
The answers from this subset are indexed, but the questions are discared after learning \modelone (i.e, they are not used for training and testing).
In the case of \manrns, IBM \modelone is trained on a subset of \compr from which we exclude QA pairs that belong to \manrns.
The split of \manr mimics the setup Surdeanu~et~al.~\shortcite{surdeanu2011learning} and the test set contains 29K queries.
Collection statistics is summarized in Table~\ref{tab:collections}.

\begin{table}[htb]
\small
\centering
\begin{tabular}{@{\hskip 0.4em}l@{\hskip 0.4em}c@{\hskip 0.4em}c@{\hskip 0.4em}c@{\hskip 0.4em}c@{\hskip 0.4em}c@{}c@{}c@{}}
\toprule
\begin{tabular}{c}Collection \\ name \\\end{tabular} & \multicolumn{5}{c}{QA pairs} & \multicolumn{1}{l}{\begin{tabular}[c]{@{}c@{}}Terms in\\ question\end{tabular}} & \multicolumn{1}{l}{\begin{tabular}[c]{@{}c@{}}Terms in\\ answer \end{tabular}} \\ 

                &   total &  train &  dev1/dev2 &  test &  tran &    &  \\ 
\midrule
\manr    & \tblfntsz 142K   & \tblfntsz 86K & \tblfntsz 7K/21K & \tblfntsz 29K  & \tblfntsz 4.2M     & \tblfntsz 13.9    & \tblfntsz 40.6                                                                                         \\
\compr   & \tblfntsz 4.4M   & \tblfntsz 212K & \tblfntsz 11K/42K & \tblfntsz 10K  & \tblfntsz 4.1M      & \tblfntsz 17.8    & \tblfntsz 34.1                                                                                         \\
\stacko  & \tblfntsz 6.2M   & \tblfntsz 298K & \tblfntsz 15K/58K & \tblfntsz 10K  & \tblfntsz 5.8M      & \tblfntsz 48.4    & \tblfntsz 33.1                                                                                         \\ \bottomrule
\end{tabular}
\caption{Collection statistics. The column \emph{tran} describes the size of the \bmmodelone training corpus. Note that for \manr
we use an \emph{external} corpus to train \bmmodelonens.}
\label{tab:collections}
\end{table}
}

\begin{figure}[tb]
\begin{center}
\includegraphics[width=0.475\textwidth]{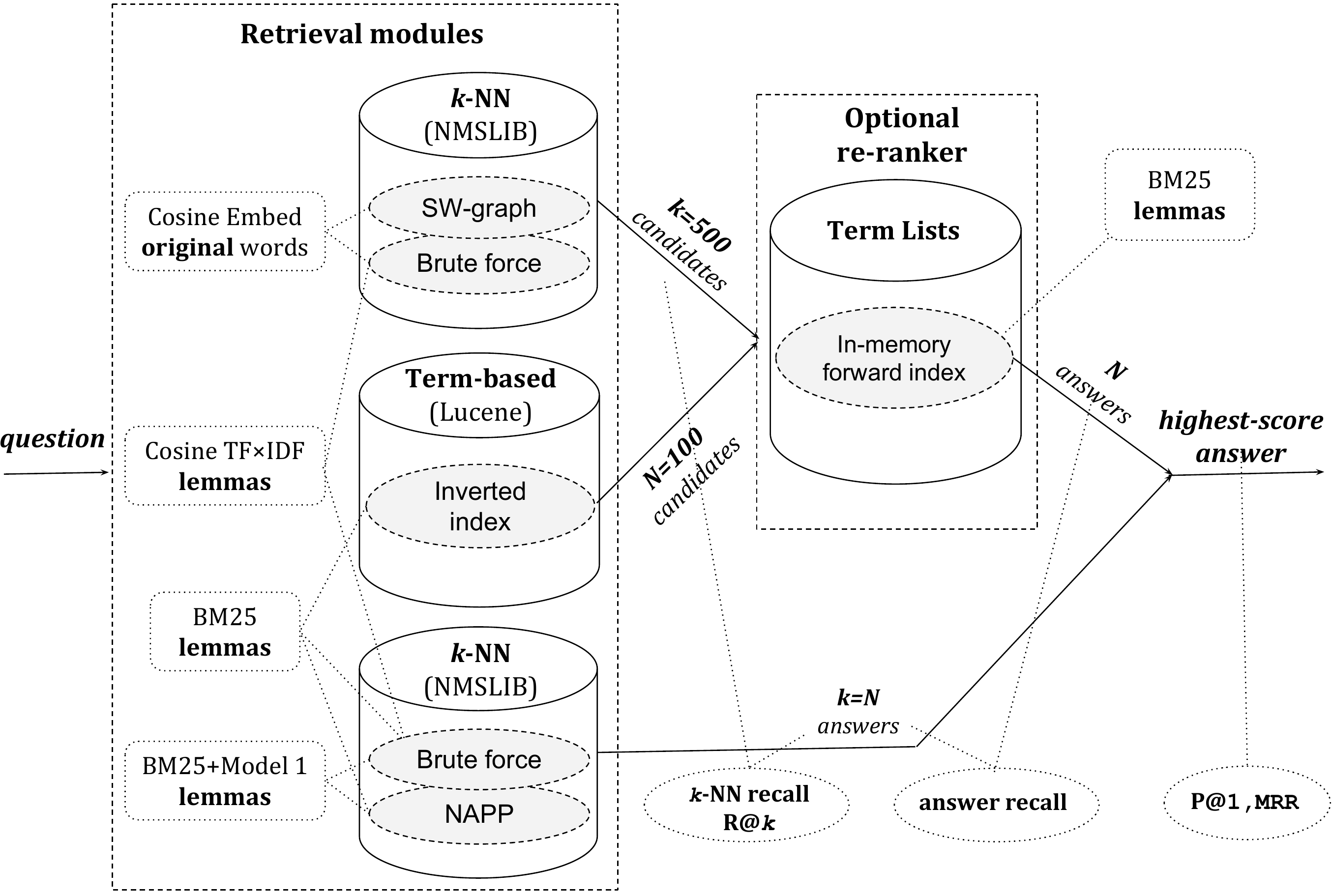}
\end{center}
\vspace*{-2.5pt}
\caption{Retrieval pipeline architecture.  We illustrate the use of evaluation metrics (inside ovals at the bottom) 
by dotted lines, which connect ovals with components for which metrics are applied.}
\vspace*{-2.5pt}
\label{fig:schema}
\end{figure}

We have implemented multiple retrieval methods and four similarity models (i.e., similarity functions). 
Retrieval methods, similarity models, and their interactions are summarized in Figure~\ref{fig:schema}.
Each retrieval method returns a ranked list of answers, which may be optionally re-ranked.
The output is a list of $N$ scored answers.
There are two classes of retrieval methods: term-based retrieval supported by Apache Lucene\footnote{\url{http://lucene.apache.org}}
and the \knn search methods implemented in the Non-Metric Space Library (NMSLIB). 
NMSLIB is an extendible framework for the \knn search in generic spaces~\cite{boytsov2013engineering}.\footnote{\url{https://github.com/searchivarius/nmslib}}
Similarity models include:
\begin{itemize}
\item  \tfidf models: the cosine similarity between \tfidf vectors (shortly \tfidfcosine)
and \bmtf  (\S \ref{sec:tfidf});
\item The cosine similarity between averaged word embeddings, henceforth, \embedcosine (\S \ref{sec:cosine_word});
\item The linear combination of BM25 and IBM \modelone scores, henceforth,
\bmmodelone (\S \ref{sec:bm25modelone}).
\end{itemize}

In the case of Lucene, we index lemmatized terms and use \bmtf as a similarity model \cite{Robertson2004}. % with parameters $k_1=1.2$ and $b=0.75$. 
We have found that Lucene's implementation of BM25 is imperfect (see \S~\ref{sec:tfidf} for details), 
which leads to at least a 10\% loss in P@1 for both \compr and  \stackons. 
To compensate for this drawback, 
we obtain 100 top-scored documents using Lucene and re-rank them using our own implementation of BM25.

In the case of NMSLIB, we use two indexing methods and the brute force search.
The indexing methods are:
the Neighborhood APProximation index (\emph{NAPP})~\shortcite{tellez2013succinct}
and the Small-World graph (\emph{SW-graph})~\shortcite{malkov2014approximate}.
They are discussed in \S~\ref{sec:knn-search}.
The SW-graph is used only with the \embedcosinens;
NAPP is applied to both \bmtf and \bmmodelonens.
We do not create an index for the \tfidfcosinens, but use the brute force search instead.
The brute force search is slow, but it is applicable to any similarity model.

Because the \tfidfcosine and \embedcosine are not very accurate,
the output from these models may be further re-ranked using BM25. 
To this end, we first retrieve 500 candidate records.
Next, we discard all but $N$ records with highest BM25 scores. 
To compare effectiveness of \tfidfcosine and \bmtfns, we also evaluate a variant where the output of \tfidfcosine is not re-ranked.

To re-rank efficiently, we use a forward index.
Given a document identifier, this index allows us to quickly retrieve the list of terms and their in-document frequencies.
Following a recent trend in high throughput in-memory database systems \cite{Kallman2008}, we load forward indices into memory.
The overall re-ranking time is negligibly small.% (less than \SI{3}{\milli\second}).
%In the remaining part of this section, we describe similarity models and \knn search methods in more detail.

Note that the depth of a candidate pool represents a reasonable efficiency-effectiveness trade-off. 
While increasing the depth of the pool improves the answer recall,
it also makes it harder to rank results accurately.
Beyond a certain point,
increasing the depth leads only to a marginal improvement in P@1 at the expense of disproportionately large computational effort. 

%\newpage
\subsection{Similarity Models}\label{sec:similmodel}
\subsubsection{Cosine \tfidf and BM25}\label{sec:tfidf}
\tfidfcosine and \bmtf are computed for \emph{lemmatized} text.
\tfidfcosine is the classic model where the similarity score is equal to the cosine similarity between \tfidf vectors \cite{Salton1986,manning2008introduction}.
An element $i$ of such a vector is equal to the product of the \emph{unnormalized} term frequency $\mbox{TF}_i$ and the inverse document
frequency (IDF). To compute IDF, we use the formula implemented in Lucene: 
\begin{equation}\label{eqIDF}
\ln \left(1 + (D - d + 0.5)/(d + 0.5)\right), 
\end{equation}
where $D$ is the number of documents and $d$ is the number of documents containing the term $i$.

BM25 scores \cite{Robertson2004} are computed as the sum of term IDFs (Eq.~\ref{eqIDF}) multiplied by respective \emph{normalized} term frequencies. 
The sum includes only terms appearing in both the query and the answer.
We also normalize BM25 scores using the sum of query term IDFs.
Normalized frequencies are as follows:
\begin{equation}\label{eqNormTF}
 \frac{
\text{TF}_i \cdot (k_1 + 1)
}
{
  \text{TF}_i + k_1 \cdot \left(1 - b + b \cdot |D|\cdot |D|^{-1}_{\text{avg}}\right),
}
\end{equation}
where $k_1$ and $b$ are parameters ($k_1=1.2$ and $b=0.75$); $|D|$ is a document length in words; $|D|_{\text{avg}}$ is the average document
length. Lucene's implementation of BM25 uses a \emph{lossy} compression for the document length, which results in reduced effectiveness.

\subsubsection{IBM Model 1}\label{sec:model1}
Computing translation probabilities via IBM Model~1 \cite{brown1993mathematics} is one common way to quantify the strength of associations
among question and answer terms. 
The transformed IBM \modelone scores are used as input to a learning-to-rank algorithm.
Specifically, we take the \emph{logarithm} of the translation probability and divide it by the number of query terms.

Let $T(q|a)$ denote a probability that a question term $q$ is a translation of an answer term $a$.
Then, a probability that a question $Q$ is a translation of an answer $A$ is equal to:
\begin{equation}\label{eq1}
\hspace{-0.5em}\begin{array}{c}
P(Q|A)=\prod\limits_{q \in Q} P(q|A) \\
P(q|A)=(1-\lambda)\left[ \sum\limits_{a \in A} T(q|a) P(a|A)\right] + \lambda P(q|C)
\\
\end{array}
\end{equation}
$T(q|a)$ is a translation probability learned by the GIZA++ toolkit \cite{Och2003} via the EM algorithm;
$P(a|A)$ is a probability that a term $a$ is generated by the answer $A$;
$P(q|C)$ is a probability that a term $q$ is generated by the entire collection $C$;
 $\lambda$ is a smoothing parameter.
$P(a|A)$ and $P(q|C)$ are computed using the maximum likelihood estimator.
For an out-of-vocabulary term $q$, $P(q|C)$ is set to a small number ($10^{-9}$). 
Similar to \bmtf and \tfidfcosinens, computation is based on the \emph{lemmatized} text.

A straightforward but slow approach to compute IBM Model~1 scores involves storing $T(q|a)$ in the form of a sparse hash table.
Then, computation of Eq.~\ref{eq1} entails one hash table lookup for each combination of question and answer terms.
We can do better by creating an inverted index for each query,
which permits retrieving query-specific entries $T(q|a)$ using the identifier of answer term $a$ as a key.
Thus, we need only one lookup per answer term.
Identifiers are indexed using an efficient hash table (the class \emph{dense\_hash\_map} from the package \emph{sparsehash})\footnote{
The code, originally written by Craig Silverstein, is now hosted at \url{https://github.com/sparsehash/sparsehash}}.
Building such an inverted index is computationally expensive (about \SI{15}{\milli\second} for each \compr
and \SI{90}{\milli\second} for each \stacko query). Yet, the cost is amortized over multiple comparisons between the query and data points.

We take several measures to maximize the effectiveness of IBM Model~1.
First, we compute translation probabilities on a symmetrized corpus as proposed  by Jeon et~al.~\shortcite{Jeon2005}.
Formally, for every pair of documents $(A, Q)$ in the parallel corpus, we expand the corpus by adding entry $(Q, A)$.

Second, unlike previous work, which seems to use \emph{complete} translation tables,
we discard all translation probabilities $T(q|a)$ below an empirically found threshold of $2.5\cdot10^{-3}$.
The rationale is that small probabilities are likely to be the result of model overfitting.
Pruning of the translation table improves both efficiency and
effectiveness. It also reduces memory requirements.

Third, following prior proposals \cite{Jeon2005,surdeanu2011learning}, 
we set $T(w|w)$, a self-translation probability, to an empirically found positive value and rescale probabilities $T(w'|w)$ so that 
$\sum_{w'} T(w'|w)=1$. 

Fourth, we make an ad hoc decision to use as
many QA pairs as possible to train IBM Model 1. A
positive impact of this decision has been confirmed
by a post hoc assessment.

%To evaluate the impact of this decision,
%we re-evaluate effectiveness of the combination IBM Model 1, simple translation features, and BM25 on \emph{dev2} set of \manrns.
%From Table~\ref{tab:model1_sizes}, we learn that achieving good performance requires a large monolingual corpus
%containing \emph{millions} of QA pairs.

Finally, we tune parameters on a development set (\emph{dev1} or \emph{dev2}).
Rather than evaluating individual performance of IBM \modelonens,
we aim to maximize performance of the model that linearly combines BM25 and IBM \modelone scores. 

\subsubsection{BM25+Model 1}\label{sec:bm25modelone}
\bmmodelone
 is a linear two-feature model, which includes BM25 and IBM \modelone scores. 
Optimal feature weights are obtained via a coordinate ascent with 10 random restarts \cite{Metzler2007}.
The model is trained via RankLib\footnote{\url{https://sourceforge.net/p/lemur/wiki/RankLib/}} using P@1 as a target optimization metric. 
To obtain training data, we retrieve $N=15$ candidates with highest BM25 scores \cite{Robertson2004} using Lucene.
If we do not retrieve a true answer, the query is discarded. 
Otherwise, we add the true answer to the training pool with the label one (which means relevant),
and the remaining retrieved answers with the label zero (which means non-relevant).

{
\begin{table}[tb]
\small
\centering
\begin{tabular}{@{}lllll@{}}\toprule
\multicolumn{5}{c}{ Prior art \cite{surdeanu2011learning}} \\ \midrule
                 &  $N=15$ & $N=25$ &  $N=50$ &  $N=100$         \\\midrule
 Recall &  0.290 &  0.328 &  0.381 &  0.434 \\ 
 P@1    &  0.499 &  0.445 &  0.385 &  0.337 \\
 MRR    &  0.642 &  0.582 &  0.512 &  0.453 \\ \midrule
\multicolumn{5}{c}{ This paper}         \\\midrule
                 &  $N=10$ & $N=17$ &  $N=36$ &  $N=72$         \\\midrule
 Recall &  0.293 &  0.331 &  0.386 &  0.438 \\
 P@1    &  0.571 (+14\%) &  0.511 (+15\%) &  0.442  (+15\%)&  0.392  (+16\%)\\
 MRR    &  0.708 (+10\%) &  0.645  (+11\%)&  0.570   (+11\%)&  0.510  (+13\%)\\ \bottomrule
\end{tabular}
\caption{Comparison of \bmmodelone against prior art on \manrns. Accuracy is computed
for several result set sizes $N$ using the methodology of Surdeanu et~al. \cite{surdeanu2011learning}. 
Each column corresponds to a different subset of queries.
\label{tab:comp_surdeanu}
}
\end{table}
}

To demonstrate that \bmmodelone delivers state of the art performance,
we compare our result on \manr against a previously published result \cite{surdeanu2011learning}.
We mimic the setup of Surdeanu et~al. \shortcite{surdeanu2011learning} and use only questions
for which a relevant answer is found (but not necessarily ranked number one).
We also split the collection in the same proportions as Surdeanu et~al. \shortcite{surdeanu2011learning}.\footnote{The exact split used by Surdeanu et al.~\cite{surdeanu2011learning} is not known. To ensure that the differences are substantial and significant, we also compute 99.9\% confidence intervals.}
Furthermore, we measure P@1 at various recall levels by varying the result sizes $N$ (which are different from those
used by Surdeanu et al \cite{surdeanu2011learning}).

According to Table~\ref{tab:comp_surdeanu}, 
our method surpasses the previously published result by 14--26\,\% in P@1,
and by 10--11\,\% in MRR despite using only two features.
This may be explained by two factors. 
First, we use a 50$\times$ larger corpus to train IBM Model~1.
Second, 
the retrieval module Terrier BM25 (employed by Surdeanu et al.~\cite{surdeanu2011learning}) seems to have inferior retrieval performance compared to Lucene. In particular, Lucene achieves a higher recall using a smaller $N$ (see Table~\ref{tab:comp_surdeanu}). 
%Our hypothesis is also supported by a recent evaluation \cite{Lin2016}.

\input plot_recall_joint.tex

\subsubsection{Cosine Embed}\label{sec:cosine_word}
Word \emph{embeddings}, also known as distributed word representations, are real-valued vectors associated with words.
Word embeddings are usually constructed in an unsupervised manner from large unstructured corpora via artificial neural networks. \cite{Collobert2011,Mikolov2013a}.
Because embeddings can capture syntactic and semantic regularities in language \cite{turney2004human,mikolov2013linguistic},
embedding-based similarity can be useful in retrieval and re-ranking tasks \cite{Fried2015,Yang2016}.
The hope here is that comparing embeddings instead of original words would help to bridge the vocabulary gap.

One popular embedding-based similarity measure is the average
cosine similarity computed for all pairs of question and answer terms.
The average pairwise cosine similarity is equal to the cosine similarity between averaged 
word embeddings of questions and answers.
In our work we use the cosine similarity between \emph{IDF-weighted} averaged embeddings. 
Here, we use embeddings of non-lemmatized terms, because this results in a slightly improved performance.
We evaluate several sets of pre-trained embeddings to select the most effective ones \cite{Mikolov2013a,pennington2014,wieting2015}.
We further improve embeddings by retrofitting \cite{faruqui2015}. In that, \modelone translation table $T(q|a)$  (see Eq.~\ref{eq1}) is used as a relational lexicon.

% Large is only used to overcome an error in ACM templates!
\subsection{Methods of {\large $k$}-NN Search}\label{sec:knn-search}
We employ a \knn retrieval framework NMSLIB,
which provides several implementations of distance
based indexing methods~\cite{boytsov2013engineering}. These indexing methods
treat data points as unstructured objects, together
with a black-box distance function. In that, the indexing and searching process exploit only values of
mutual object distances.
NMSLIB can be extended by implementing new
black-box ``distance'' functions. In particular, we
add an implementation for the similarity functions \bmtfns, \tfidfcosinens, \embedcosinens, and \bmmodelone (see in \S \ref{sec:similmodel}). 
None of these similarity functions is a metric distance.
In particular, in the case of \bmmodelone the ``distance'' lacks symmetry.

Because exact \knn search is too slow to be practical, we resort to an \emph{approximate} procedure,
which does not necessarily find all $k$ nearest neighbors.
The accuracy of the \knn search is measured using a recall denoted as R@k.
 R@k is equal to the fraction of true $k$-nearest neighbors found. 

NMSLIB reads contents of the forward index
(created by a separate indexing pipeline) into memory and
builds an additional \emph{in-memory} index. 
In this work, we create indices using one of the following method:
the Neighborhood APProximation index (\emph{NAPP}) due to Tellez et al.~\shortcite{tellez2013succinct}
or the proximity graph method called a Small-World graph (\emph{SW-graph}) due to Malkov~et~al.~\shortcite{malkov2014approximate}.

NAPP is a \emph{pivoting} method that arranges points based on their distances to pivots.
This is a filtering method: Candidate points share \ttt{numPivotSearch} closest pivots with the query point.
The search algorithm employs an inverted index. Unlike term-based indices,
however, for each pivot the index keeps references to close data points.
More specifically, the pivot should be one of the point's \ttt{numPivotIndex} closest pivots.
Answering a query requires efficient merging of posting lists. 
Merging of posting lists represents a substantial overhead.

Tellez et al.~\shortcite{tellez2013succinct} use pivots randomly sampled from the data set,
but we find that for \emph{sparse} data such as \tfidf vectors
substantially shorter retrieval times---sometimes by orders of magnitude---can be obtained by using a special pivot generation algorithm.
Specifically, pivots are generated as pseudo-documents containing $K$ entries
sampled from the set of $M$ most frequent words (in our setup $K=1000$ and $M=50000$).
A more detailed description and analysis of this approach will be presented elsewhere.

During indexing, we have to compute the distances between a data point and every pivot.
Because there are thousands of pivots, this operation is quite expensive, especially for \bmmodelonens.
To optimize computation of Eq.~\ref{eq1}, we organize all pivot-specific $T(q|a)$ entries in the form of the inverted index. 

A proximity graph is a data structure, where data points are nodes.
Sufficiently close nodes, i.e., \emph{neighbors}, are connected by edges.
Searching starts from some, e.g., random, point/node and traverses the graph until it stops discovering new points sufficiently close to the query 
or after visiting a given number of nodes.
\cite{arya1993approximate,sebastian2002metric,houle2005fast,hajebi2011fast,dong2011efficient,wang2015fast}.
Specifically, the SW-graph algorithm (implemented in NMSLIB) keeps a list of \ttt{efSearch} points sorted in
the order of increasing distance from the query as well as a candidate queue.
Traversal proceeds in the best-first manner, by exploring the neighborhood of the candidate that is closest to the query.
If a candidate neighbor is closer to the query than the \ttt{efSearch}-\textit{th} closest point seen so far,
it is added to the candidate queue. Otherwise, the neighbor is discarded. 
The traversal stops when the candidate queue is exhausted.  
To improve recall, the algorithm may restart several times.
For our data, however, it is more efficient to start from a single point and search using a large-enough value of \ttt{efSearch}.

SW-graph works well for dense vectorial data (i.e., embeddings), 
where it outstrips NAPP by an order of magnitude. 
SW-graph was found to be much faster \cite{NaidanBN15}
than the multi-probe LSH due to Dong et al.~\cite{Dong2008}.
In a public evaluation in May 2016,\footnote{\url{https://github.com/erikbern/ann-benchmarks}} SW-graph outperformed  
two efficient popular libraries:
FLANN \cite{muja2014scalable} and Annoy\footnote{\url{https://github.com/spotify/annoy}}.
SW-graph was also mostly faster than a novel LSH algorithm \cite{NIPS2015_5893}.
In contrast, NAPP substantially outperforms SW-graph for sparse \tfidf data, 
i.e., for models \bmtf and \bmmodelonens.

%% file: plot_recall_joint.tex
\begin{figure*}[htb]\centering
\subfloat[\stackons]{\label{fig:recall_stack} 
\includegraphics[width=0.48\textwidth]{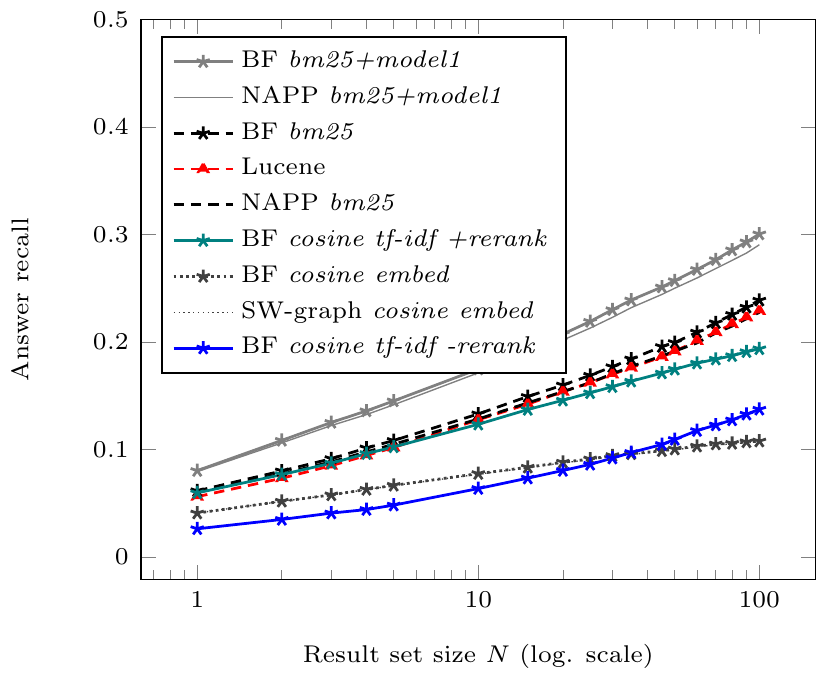}
}
\subfloat[\comprns]{\label{fig:recall_compr} 
\includegraphics[width=0.48\textwidth]{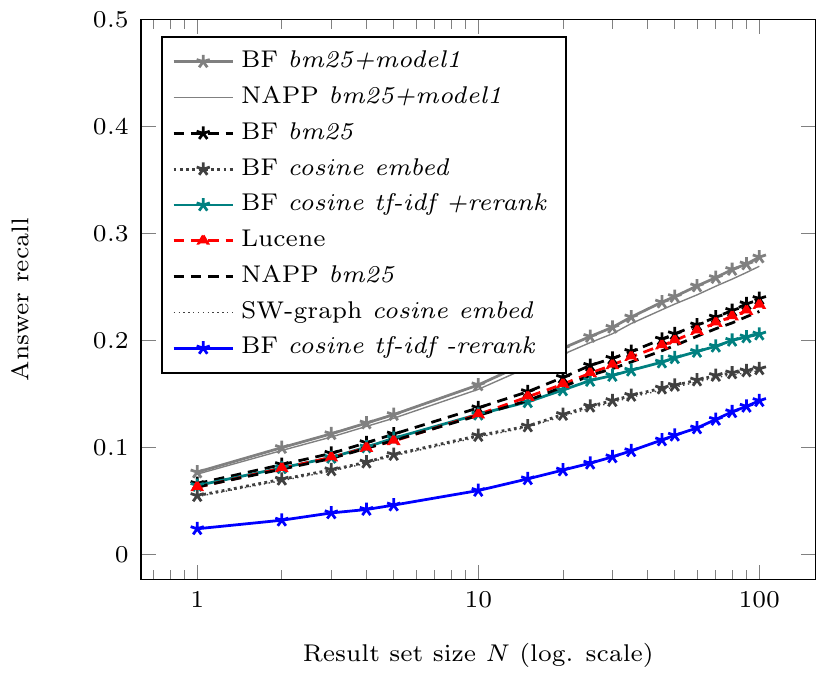}
}
\caption{\label{fig:recall_all}
The answer recall at different $N$. 
(BF stands for brute-force; \emph{+rerank} and \emph{-rerank} indicate if an optional re-ranker is used).
}
\end{figure*}

%% file: main_exper.tex
\input main_table.tex

Experiments are carried out on Amazon EC2 instance \emph{r3.4xlarge}, which has 16 virtual cores and 122 GB of memory.
The main retrieval pipeline, which is implemented in Java (1.8.0\_11), uses 16 search threads.
We use a modified version of NMSLIB 1.5,\footnote{\url{https://github.com/searchivarius/nmslib/tree/nmslib4a_cikm2016}}
 which operates as a server processing queries via TCP/IP.
NMSLIB and its extensions are written in C++ and compiled using GCC 4.8.4 with optimization flags \ttt{-O3} and \ttt{-march=native}. 
%The cosine similarity between embeddings is computed by an optimized function that uses SIMD instructions.
%NAPP does not use any low-level optimizations.
Lucene version is 4.10.3. The retrieval architecture (see \S~\ref{sec:approach}) is outlined in Figure \ref{fig:schema}.

The collection processing/indexing system is implemented in Java. 
It employs the framework Apache UIMA
and UIMA components from DKPro Core \cite{dkpro2014}.\footnote{\url{https://dkpro.github.io/dkpro-core/}}
Translation probabilities are computed using GIZA++ toolkit \cite{Och2003} via the EM algorithm (five iterations).\footnote{\url{https://github.com/moses-smt/giza-pp}}

For each \emph{approximate} \knn pipeline, we execute several runs with different parameters. 
In the case of SW-graph, we vary the parameter \ttt{efSearch}. In the case of NAPP, we 
vary the number of indexed pivots (parameter \ttt{numPivotIndex}) and the number of pivots
that should be shared between the query and an answer (\ttt{numPivotSearch}).
Optimal parameters have been found on a \emph{dev1} set (using a subset of 5K queries).

Retrieval times are measured by a special client application that submits search requests to either Lucene or NMSLIB.
In the case of Lucene, we ``warm up'' the index by executing the whole set of queries twice. 
Run-times are measured only for the third run.

Effectiveness of \emph{retrieval runs} is measured using an external application,
namely, {trec\_eval} 9.0.4.\footnote{\url{https://github.com/usnistgov/trec_eval}}
The main experimental results are presented in Figure~\ref{fig:recall_all} and Table~\ref{tab:main}.
For Table~\ref{tab:main} we compute statistical significance of results using the t-test with a subsequent Bonferroni adjustment for multiple testing.
This adjustment for multiple testing consists in multiplying p-values by the total number of runs for $N=100$ (to save space, some of the runs are not shown in the table). The significance level is 0.01.

In Figure~\ref{fig:recall_all} we plot the answer recall (measured for a set of top-$N$ ranked documents) for nine implemented retrieval modules. 
Note that approximate \knn methods are represented by their most accurate runs.
Exact brute force \knn runs are plotted using thicker lines (with star marks) of the same style/color as corresponding approximate runs.
Their mnemonic names start with the word BF (short for brute force).

The cosine-similarity models are the least effective. The recall of the brute force run {BF \tfidfcosine \emph{-rerank}} 
is less than half of that for the brute force run BF \bmtfns.
We can nearly match the performance of \bmtf by adding a BM25-based optional re-ranker (the run BF \tfidfcosine \emph{+rerank}).
In contrast, the cosine-similarity between averaged word embeddings (e.g., the run BF \embedcosinens) is much worse than \bmtf
despite using the re-ranker!
Somewhat surprisingly, the cosine similarity between \tfidf vectors \emph{without} re-ranker is sometimes more effective 
than the cosine similarity between word embeddings whose performance is boosted by the re-ranker (see Panel \ref{fig:recall_stack}
in Figure \ref{fig:recall_all}). 
This is a discouraging finding given that embedding-based retrieval can be quite efficient (see Table~\ref{tab:main}). 
It remains to be verified if better results can be obtained with document embeddings
that compute vectorial representations of complete sentences or even documents \cite{iyyer2014,le2014distributed}.

Note that all BM25-based runs have similar performance. However, \bmmodelone has a recall that is 16\% higher in the case of \compr and 26\% higher
in the case of \stackons.
For \bmtfns, it is possible to match the recall of \bmmodelone by increasing $N$. 
However, this may increase a load on a downstream re-ranking module.
For example, in the case of \stackons,
\bmmodelone has a nearly 0.2 answer recall for $N=10$ (Panel \ref{fig:recall_stack} in Figure \ref{fig:recall_all}).
To obtain the same recall level using Lucene, we need to use $N>20$.

Next, we compare efficiency of \knn search methods against that of Lucene. 
Note that Lucene is a strong baseline, which fares well against optimized C++ code, especially for disjunctive queries \cite{Vigna2013}.
Lucene's average retrieval times are equal to \ms{80} for \compr and \ms{620} for \stacko (see Table~\ref{tab:main}).
There are at least two factors that contribute to the difference in retrieval times between two collections:
(1) questions in \stacko have 2.7$\times$ as many terms, (2) \stacko has 1.4$\times$ as many answers
(see Table~\ref{tab:collections}).

SW-graph is quite fast for both collections. For example, for \stackons, it can answer queries in \ms{340} at the expense of
losing only 1.3\% answers compared to the brute force search (As it is recently reported by Malkov and Yashunin \cite{Malkov2016},
an improved, hierarchical, variant of SW-graph is even more accurate and/or efficient). In other words, the approximate search is nearly as accurate
as the exact one. This is why in Figure~\ref{fig:recall_all} the best \emph{approximate} SW-graph run for the model \embedcosine and the run BF \embedcosine are
hard to distinguish.
However, the model \embedcosine is not very effective. It does not bridge the vocabulary gap and is even worse than \tfidfcosinens.

In the case of \bmtfns, NAPP works well for \stackons, but not for \comprns. For example, in the case of \stackons,
it answers queries in \ms{230} while losing only 2.1\% in P@1 and 4.7\% in the answer recall. 
This is nearly 2.7$\times$ faster than Lucene and 15$\times$ faster than the brute force search using \bmtfns.

For the more complicated model \bmmodelonens, NAPP delivers similar speed ups over the brute force search
for both collections. 
However, it is always slower than Lucene in the case of \comprns.
In the case of \stackons, NAPP is up to 1.5$\times$ faster than Lucene. For the fastest posted retrieval time of \ms{400}  
it delivers P@1 equal to 0.074 and the answer recall equal to 0.252.

Despite some degradation in comparison to the corresponding exact brute force run,
this represents an impressive 19.3\% improvement in P@1 and 5.4\% improvement in the answer recall compared
to the brute force \bmtfns. 
The second slowest \bmmodelone run obtained by NAPP is nearly as efficient as Lucene,
but it outperforms \bmtf by 27.4\% in P@1 and by 18.4\% in recall.
%These differences are statistically significant.
%Also note that here we do not compare effectiveness against Lucene, which has somewhat worse P@1 and answer recall
%compared to BF \bmtf.

% \usepackage{booktabs}
\begin{table}[bt]
\small
\centering

\begin{tabular}{@{}cccc@{}}
\toprule
\multicolumn{4}{c}{\compr}                                                                                                                                                          \\ \midrule
\multicolumn{2}{c}{\bmmodelone}                                                                     & \multicolumn{2}{|c}{\bmtf}                                                    \\ \midrule
R@1   & \multicolumn{1}{c|}{\begin{tabular}[c]{@{}c@{}}Reduction in \\ distance comp.\end{tabular}} & R@1   & \begin{tabular}[c]{@{}c@{}}Reduction in\\ distance comp.\end{tabular} \\ \midrule
0.982 & \multicolumn{1}{c|}{8.7}                                                                    & 0.982 & 3.7                                                                   \\
0.968 & \multicolumn{1}{c|}{61}                                                                     & 0.970 & 20                                                                    \\
0.961 & \multicolumn{1}{c|}{142}                                                                    & 0.963 & 48                                                                    \\
0.952 & \multicolumn{1}{c|}{246}                                                                    & 0.956 & 98                                                                    \\
0.930 & \multicolumn{1}{c|}{434}                                                                    & 0.927 & 226                                                                   \\ \midrule
\multicolumn{4}{c}{\stacko}                                                                                                                                                         \\ \midrule
\multicolumn{2}{c}{\bmmodelone}                                                                     & \multicolumn{2}{|c}{\bmtf}                                                    \\ \midrule
R@1   & \multicolumn{1}{c|}{\begin{tabular}[c]{@{}c@{}}Reduction in\\  distance comp.\end{tabular}} & R@1   & \begin{tabular}[c]{@{}c@{}}Reduction in\\ distance comp.\end{tabular} \\\midrule
0.982 & \multicolumn{1}{c|}{13.4}                                                                   & 0.980 & 157                                                                   \\
0.970 & \multicolumn{1}{c|}{39}                                                                     & 0.972 & 208                                                                   \\
0.964 & \multicolumn{1}{c|}{64}                                                                     & 0.964 & 260                                                                   \\
0.957 & \multicolumn{1}{c|}{97}                                                                     & 0.959 & 287                                                                   \\
0.948 & \multicolumn{1}{c|}{137}                                                                    & 0.955 & 315                                                                   \\ \bottomrule
\end{tabular}
\caption{
Reduction in the number of the distance computation for two similarity models
at approximately equal levels of R@1 (larger reduction is better).
Using 5K queries from \emph{dev1} set.
}
\label{tab:bm25_vs_exper1}
\end{table}

Also note that for \comprns, NAPP \bmmodelone can be both faster and more accurate than NAPP \bmtfns. 
This is quite surprising given that \bmmodelone is expensive to compute. 
Specifically, the corresponding
brute force run is nearly 5$\times$ slower compared to the brute force run of \bmtfns.
There are two reasons for why NAPP \bmmodelone can be more faster and accurate than NAPP \bmtfns. First, there is a high overhead related to merging pivot posting lists.
By varying method's parameters we can reduce the amount of time spent on computation of 
the distance (at the expense of search accuracy).
At some point the time spent on distance computation 
becomes so small so that the overhead related to processing of posting lists starts to dominate the overall time.

Second, there are differences in filtering effectiveness of the methods.
To demonstrate this, we evaluate 
 the reduction in the number of distance computations compared to the brute force search.
For example, if an algorithm answers a query by checking only 10\%
of data points, the reduction in the number of distance computations is 10.
Reductions in the number of distance computations are compared for nearly equal
values of the \knn recall R@1,
which is equal to the fraction of true nearest neighbors
found by the retrieval module (R@1 should not be confused with the answer recall).
The results of this comparison are presented in Table~\ref{tab:bm25_vs_exper1}.
For NAPP, the more pivots are indexed, the fewer distance computations are necessary to achieve a given accuracy level.
Thus, to make a fair comparison, we index equal number of pivots for both \bmtf and \bmmodelonens.

According to Table~\ref{tab:bm25_vs_exper1}, 
in the case of \comprns, it takes 2-3$\times$ fewer distance computations for the model \bmmodelone than for \bmtfns.
In contrast, in the case of \stackons, answering queries for the model \bmtf takes significantly fewer distance computations
than for \bmmodelonens.
Furthermore, the reduction in the number of distance computations for \bmtf on \stacko can be two orders of magnitude
higher compared to that of \comprns. What are the possible explanations for these stark differences?

We think that pivoting methods are effective only if comparing a query and an answer with the same pivot
provides a meaningful information regarding their proximity.
In the case of a simple \bmtf model, 
this is only possible if the pivot, the query, and the answer have at least one common term. 
Such an overlap is much more likely in the case of \stacko where questions are nearly 3$\times$ longer compared to \comprns.
In contrast, for the model \bmmodelone information regarding proximity of answers and queries 
may be obtained if pivots, queries, and answers share
only related but not necessarily identical terms. 
Thus, using \bmmodelone is more advantageous in the case of short queries (e.g., in the case of \comprns).

\begin{figure*} [htb]\centering
\subfloat[\label{ProxComprBM25Model1}\comprns: \bmmodelonens]{\includegraphics[width=0.38\textwidth]{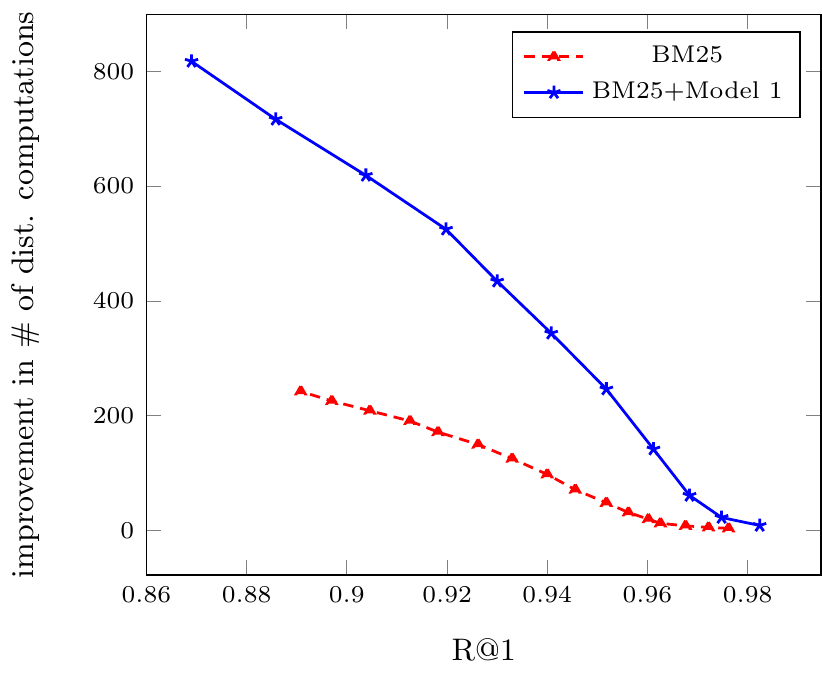}}
\hspace{2em}
\subfloat[\label{ProxComprBM25}\comprns: \bmtfns]{\includegraphics[width=0.38\textwidth]{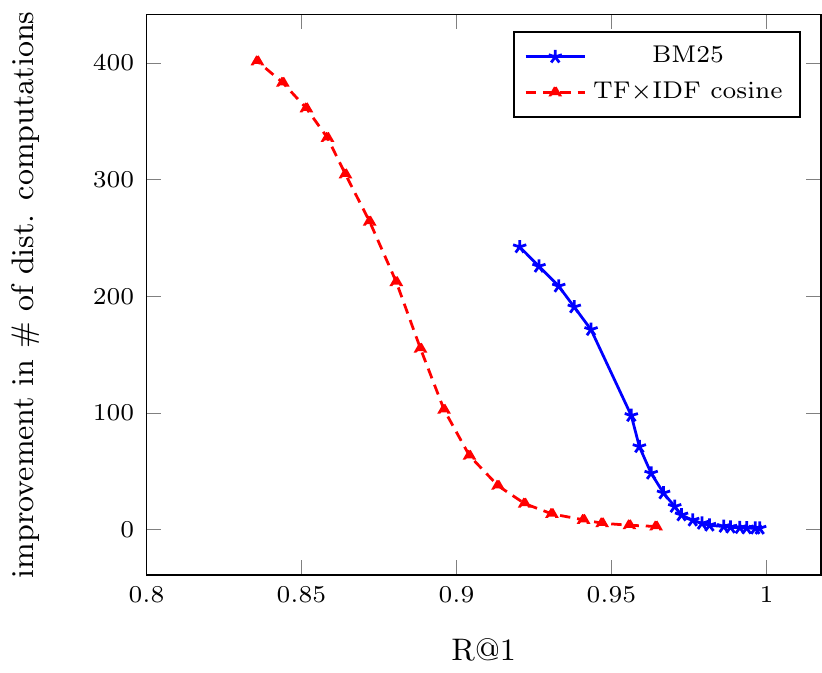}}
\\
\subfloat[\label{ProxStackBM25Model1}\stacko: \bmmodelonens]{\includegraphics[width=0.38\textwidth]{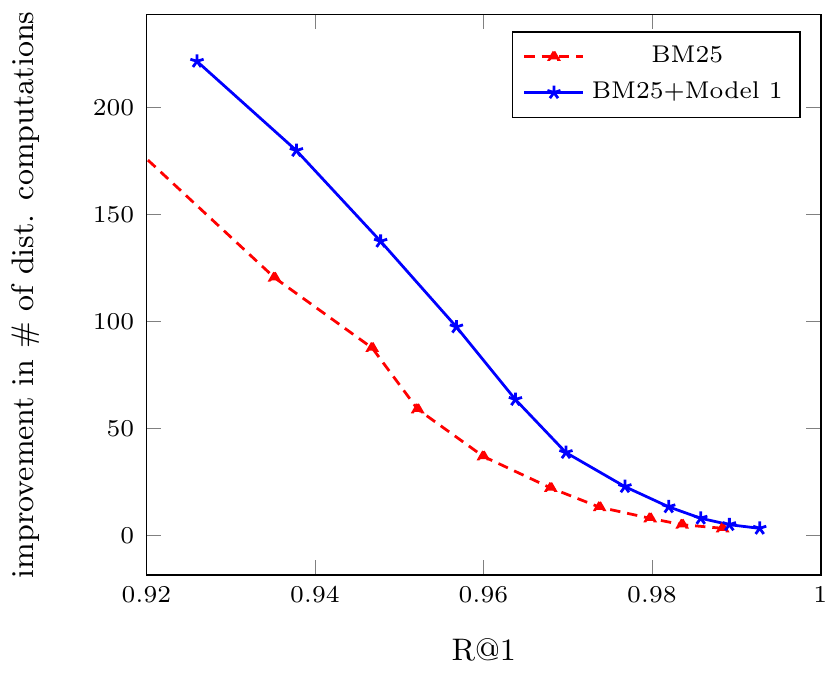}}
\hspace{2em}
\subfloat[\label{ProxStackBM25}\stacko: \bmtfns]{\includegraphics[width=0.38\textwidth]{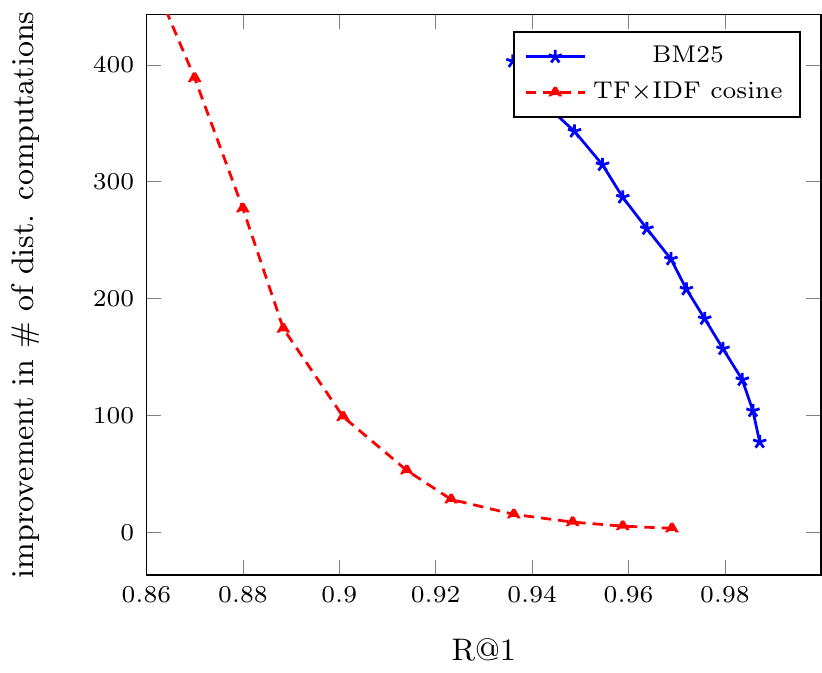}}
\caption{
Filtering effectiveness of NAPP for original and a proxy distance function (when computing distances to pivots). 
The curves for the original distance are blue while proxy distance curves are red.
Filtering effectiveness is measured using via reduction in the number of distance computations (larger is better).
The left column has data for the distance \bmmodelone and the right column has data for \bmtfns. 
Using 5K queries from \emph{dev1} set.
\label{AllPlotsProxyDistance}
}
\end{figure*}

{
%\color{blue}
To further illustrate importance of using the right function to compute distance to pivots,
we evaluate filtering effectivness in two scenarious: (1) when the distance to pivots is computed
using an original distance function and (2) when the distance to pivots is computed using a different, i.e., proxy function.
For each scenarios, we use two models: \bmmodelone and \bmtfns.
In the case of \bmmodelonens, the proxy distance is \bmtfns. In the case of \bmtfns, the proxy distance is \tfidfcosinens.
The results are presented in Figure~\ref{AllPlotsProxyDistance} where the curves corresponding to the original distance are blue
and the curves corresponding to the proxy distance are red.

Panels \ref{ProxComprBM25Model1} and \ref{ProxStackBM25Model1} show us what happens if the distance to pivots is computed
using cheap \bmtf instead of expensive \bmmodelonens.
We can see that resorting to using the proxy distance makes us check more candidate documents to achieve the same level of recall.
In other words relying on the proxy distance has a negative effect on filtering effectiveness. 
In turn, this can drastically reduce overall search efficiency.

The difference is bigger for Panel  \ref{ProxComprBM25Model1}, which corresponds to the collection \comprns.
A likely explanation of this difference is the above-described disparity in query lengths between two collections.
In the case of \stacko queries are long and there is a bigger overlap between queries and answer documents. 
This is why the similarity function that relies on a pure lexical match (in this case \bmtfns) allows us
to find answers rather effectively. In the case of \compr a lexical overlap between queries and answer documents
is less likely, which can be, nevertheless, remedied by enhancing \bmtf model with \modelone scores.
However, when \modelone scores are excluded---by using the proxy distance function to compute distance to pivots---this
exclusion has a larger negative effect for \compr than for \stackons.

Panels \ref{ProxComprBM25} and \ref{ProxStackBM25} show us what happens if the distance to pivots is computed
using \tfidfcosine instead of \bmtfns. In this case, such a replacement leads to a much 
larger performance deterioration than removal of \modelone scores.
This is not surprising: As we can see from Figure~\ref{fig:recall_all}, 
there is a much bigger gap in effectiveness between \bmtf and \tfidfcosine than between \bmmodelone and \bmtfns. 
Thus, replacing \bmtf with \tfidfcosine has also a larger negative effective on filtering effectiveness (than
replacing \bmmodelone with \bmtfns).
}

%As a reminder, effectiveness of the candidate generator is measured using a candidate recall.
%If candidates are generated using the \knn search, we \emph{additionally} measure the \knn recall
% R@k, which is equal to the fraction of true $k$-nearest neighbors found. 
%These two effectiveness metrics should not be confused.

%% file: main_table.tex
{
%\renewcommand{\arraystretch}{0.9}
% \usepackage{booktabs}
\begin{table*}[tb]
%\tblfntsz
\small
\centering
\begin{tabular}{@{}cclclccclclc@{}}
\toprule
\multicolumn{6}{c}{\textbf{\stacko}} & \multicolumn{6}{c}{\textbf{\compr}} \\ \midrule
\begin{tabular}[c]{@{}c@{}}Query\\ time \end{tabular} & \begin{tabular}[c]{@{}c@{}}Speed-up\\ (over BF)\end{tabular} & P@1 & \begin{tabular}[c]{@{}c@{}}P@1\\ loss\end{tabular} & \begin{tabular}[c]{@{}c@{}}Answer\\ recall\end{tabular} & \multicolumn{1}{c|}{\begin{tabular}[c]{@{}c@{}}Answer recall\\ loss\end{tabular}} & \begin{tabular}[c]{@{}c@{}}Query \\ time \end{tabular} & \begin{tabular}[c]{@{}c@{}}Speed-up\\ (over BF)\end{tabular} & P@1 & \begin{tabular}[c]{@{}c@{}}P@1\\ loss\end{tabular} & \begin{tabular}[c]{@{}c@{}}Answer\\ recall\end{tabular} & \begin{tabular}[c]{@{}c@{}}Answer recall\\ loss\end{tabular} \\ \midrule
\multicolumn{12}{c}{\textbf{BF \bmmodelone}} \\ \midrule
\second{20.5} &  & 0.081* &  & 0.300* & \multicolumn{1}{c|}{} & \second{8} &  & 0.077* &  & 0.278* &  \\ \midrule
\multicolumn{12}{c}{\textbf{NAPP \bmmodelone}} \\ \midrule
\second{0.84} & 24 & 0.080* & 1.1\% & 0.290* & \multicolumn{1}{c|}{3.3\%} & \second{0.70} & 11 & 0.075* & 2.6\% & 0.269* & 3.1\% \\
%716 & 29 & 0.079* & 1.5\% & 0.287* & \multicolumn{1}{c|}{4.5\%} & 385 & 21 & 0.074 & 3.9\% & 0.265 & 4.8\% \\
\second{0.65} & 32 & 0.079* & 2.2\% & 0.283* & \multicolumn{1}{c|}{5.7\%} & \second{0.30} & 27 & 0.074* & 4.7\% & 0.261* & 6.1\% \\
%611 & 34 & 0.079* & 2.2\% & 0.280* & \multicolumn{1}{c|}{7.0\%} & 269 & 30 & 0.072 & 6.3\% & 0.257 & 7.4\% \\
\second{0.50} & 35 & 0.078* & 3.1\% & 0.276* & \multicolumn{1}{c|}{8.2\%} & \second{0.21} & 38 & 0.073 & 5.7\% & 0.256* & 8.1\% \\
%510 & 40 & 0.077* & 4.8\% & 0.272* & \multicolumn{1}{c|}{9.4\%} & 188 & 42 & 0.071 & 7.8\% & 0.248 & 10.6\% \\
\second{0.47} & 44 & 0.076* & 5.8\% & 0.263* & \multicolumn{1}{c|}{12.4\%} & \second{0.18} & 45 & 0.070 & 9.7\% & 0.234 & 16.0\% \\
%449 & 46 & 0.074* & 7.6\% & 0.254* & \multicolumn{1}{c|}{15.5\%} & 159 & 50 & 0.070 & 8.9\% & 0.246 & 11.4\% \\
\second{0.40} & 52 & 0.074* & 8.2\% & 0.252* & \multicolumn{1}{c|}{16.1\%} & \second{0.09} & 89 & 0.068 & 12.4\% & 0.227* & 18.5\% \\ \midrule
\multicolumn{12}{c}{\textbf{BF \bmtf}} \\ \midrule
\second{3.5} &  & 0.062 &  & 0.239 & \multicolumn{1}{c|}{} & \second{1.7} &  & 0.067 &  & 0.239 &  \\ \midrule
\multicolumn{12}{c}{\textbf{NAPP \bmtf}} \\ \midrule
\second{0.23} & 15 & 0.060 & 2.1\% & 0.228* & \multicolumn{1}{c|}{4.7\%} & \second{0.37} & 5 & 0.063* & 5.1\% & 0.227* & 5.0\% \\
%223 & 16 & 0.060 & 2.1\% & 0.226 & \multicolumn{1}{c|}{5.5\%} & 229 & 7 & 0.063* & 5.4\% & 0.224* & 6.3\% \\
\second{0.14} & 25 & 0.059* & 5.0\% & 0.221* & \multicolumn{1}{c|}{7.6\%} & \second{0.18} & 9 & 0.062* & 6.6\% & 0.222* & 7.1\% \\
%86 & 41 & 0.058* & 6.3\% & 0.214 & \multicolumn{1}{c|}{10.4\%} & 162 & 10 & 0.062* & 7.2\% & 0.220* & 7.9\% \\
\second{0.07} & 50 & 0.057* & 7.6\% & 0.208* & \multicolumn{1}{c|}{12.9\%} & \second{0.15} & 11 & 0.061* & 8.4\% & 0.217* & 9.2\% \\
%65 & 54 & 0.056 & 9.6\% & 0.201 & \multicolumn{1}{c|}{15.8\%} & 151 & 11 & 0.061* & 8.9\% & 0.216* & 9.9\% \\
\second{0.06} & 56 & 0.055* & 11.2\% & 0.195* & \multicolumn{1}{c|}{18.5\%} & \second{0.15} & 11 & 0.060* & 9.3\% & 0.212* & 11.2\% \\
%62 & 57 & 0.054* & 11.8\% & 0.187* & \multicolumn{1}{c|}{21.8\%} & 147 & 12 & 0.060* & 10.7\% & 0.207* & 13.6\% \\ 
\midrule
\multicolumn{12}{c}{\textbf{Lucene BM25}} \\ \midrule
\second{0.62} &  & 0.062 &  & 0.229* & \multicolumn{1}{c|}{} & \second{0.08} &  & 0.067 &  & 0.233* &  \\ \midrule
%\multicolumn{12}{c}{\textbf{Lucene (with query expansion)}} \\ \midrule
%\second{1.6} &  & 0.062 &  & 0.228 & \multicolumn{1}{c|}{} & \second{0.23} &  & 0.067 &  & 0.226* &  \\ \midrule

%\multicolumn{12}{c}{\textbf{BF \tfidfcosine}} \\ \midrule
%4004 &  & 0.060 &  & 0.194 & \multicolumn{1}{c|}{} & 2029 &  & 0.065 &  & 0.206 &  \\ \midrule
\multicolumn{12}{c}{\textbf{BF \embedcosine}} \\ \midrule
\second{3.9} &  & 0.041* &  & 0.108* & \multicolumn{1}{c|}{} & \second{2.7} &  & 0.055* &  & 0.174* &  \\ \midrule
\multicolumn{12}{c}{\textbf{SW-graph \embedcosine}} \\ \midrule
\second{0.78} & 5 & 0.041* & -0.2\% & 0.107* & \multicolumn{1}{c|}{0.6\%} & \second{0.19} & 14 & 0.054* & 1.6\% & 0.172* & 1.0\% \\
%519 & 7 & 0.041* & 0.5\% & 0.107 & \multicolumn{1}{c|}{0.8\%} & 115 & 23 & 0.054* & 2.5\% & 0.171* & 1.4\% \\
\second{0.40} & 10 & 0.041* & 0.5\% & 0.107* & \multicolumn{1}{c|}{0.8\%} & \second{0.09} & 31 & 0.054* & 3.1\% & 0.170* & 1.9\% \\
%374 & 10 & 0.041 & 0.5\% & 0.107 & \multicolumn{1}{c|}{0.9\%} & 79 & 34 & 0.054* & 3.3\% & 0.170* & 2.1\% \\
\second{0.34} & 11 & 0.041* & 0.7\% & 0.106* & \multicolumn{1}{c|}{1.3\%} & \second{0.07} & 37 & 0.053* & 3.6\% & 0.170* & 2.4\% \\
%215 & 18 & 0.040 & 2.9\% & 0.104 & \multicolumn{1}{c|}{3.2\%} & 43 & 62 & 0.051* & 7.4\% & 0.166* & 4.6\% \\
\second{0.13} & 29 & 0.039* & 4.9\% & 0.102* & \multicolumn{1}{c|}{5.8\%} & \second{0.03} & 104 & 0.050* & 10.3\% & 0.160* & 7.8\% \\ 
\midrule
%\multicolumn{12}{c}{\textbf{BF \tfidfcosine without re-ranking}} \\ \midrule
%4008 &  & 0.026* &  & 0.137* &  & 2032 &  & 0.025 &  & 0.144 &  \\ \bottomrule
\end{tabular}
\caption{Efficiency-effectiveness trade-offs of retrieval modules for $N=100$ (brute force \tfidfcosine runs are omitted).
Statistically significant differences (at level 0.01) from \textbf{BF \bmtfns} are marked with *. 
P-values are adjusted for multiple testing via the Bonferroni correction.
\label{tab:main}
}
\end{table*}
}

%% file: discussion.tex
The \knn search is an extensively studied area.
For a detailed discussion the reader is addressed to the surveys of metric 
\cite{chavez2001searching} and non-metric \cite{skopal2006fast} access methods,
as well to the recent survey of hashing techniques \cite{wang2014hashing}.\footnote{In addition to the \knn search, hashing techniques are often used for
near-duplicate detection \cite{manning2008introduction}, 
which consists in finding objects with a high degree of similarity.
It is a related but distinct problem, which is often solved
by applying high-precision low-recall techniques \cite{manning2008introduction}.
However, these techniques are not applicable to broader search tasks such
as finding answers relevant to a given question.}

The \knn search is a popular technique in IR and NLP, where the
following two approaches are typically used.
The first approach relies on a term-based inverted index
in retrieving documents that share common terms with the query.
These documents are further re-ranked using some similarity function.
Dynamic and static pruning can be used to improve efficiency, sometimes at the expense
of decreased recall \cite{Turtle1995,Carmel2001,Ding2011}.
This approach supports arbitrary similarity functions,
but it suffers from the problem of the vocabulary mismatch \cite{Berger2000,surdeanu2011learning,Fried2015}.

The second approach involves carrying out the \knn search via LSH \cite{petrovic2010streaming,Ture2011,Moran2016,Li2014}.
It is most appropriate for the cosine similarity. 
For example, Li et al.~\cite{Li2014} propose the following two-stage scheme
to the task of finding thematically similar documents. In the first step
they retrieve candidates using LSH. Next, these candidate are re-ranked
using the Hamming distance between quantized \tfidf vectors.
Li et al.~\cite{Li2014} find that their approach is up to 30$\times$ faster
than the classic term-based index while sometimes being equally accurate.

Petrovi{\'c} et al.~\cite{petrovic2010streaming} applied a hybrid
of LSH and the term-based index to the task of the streaming First Story Detection (FSD).
The LSH keeps a large chunk of sufficiently recent documents,
while the term-based index keeps a small subset of recently added documents.
They report their system to be substantially faster than the 
state-of-the-art system---which relies on the classic term-based index---while being similarly effective.
In a follow up work, 
Petrovi{\'c} et al.  \cite{Petrovic2012} incorporate term
associations into the similarity function.
Their solution relies on an approximation for the kernelized cosine similarity.
The associations are obtained from an external paraphrasing database.
Moran et al.~\cite{Moran2016} use the same method as Petrovi{\'c} et al.  \cite{Petrovic2012}, 
but find synonyms via the \knn search in the space of word embeddings (which works better for Twitter data).
Moran et al.~\cite{Moran2016} as well as Petrovi{\'c} et al. \cite{Petrovic2012}
calculate performance using an aggregated metric designed specifically for the FSD task.
Unfortunately, they do not report performance gains using standard IR metrics such
as precision and recall.

Most importantly, as shown in the literature (see \cite{Whissell2013} and references therein), 
similarity functions based on the cosine similarity are not especially effective.
In particular, compared to \bmtfns,
our implementation of the \tfidf cosine similarity finds 2$\times$ \emph{fewer}
answers for any given rank $N$(see Figure~\ref{fig:recall_all}). 
It is possible to improve answer recall by increasing $N$.
However, this has effect on search module' performance.
In particular, if we retrieve top-$N$ entries using an approximate \knn
algorithm, as $N$ increases, accuracy or the efficiency of the search decreases.
Simply speaking, it is easier to carry out an accurate 1-NN search than
an accurate 500-NN search. 
%This observation comes from authors' personal experience with the \knn search.

One notable exception is a recent paper by Brokos et al.~\cite{brok2016} who, in contrast to our findings, 
learned that the cosine-similarity between averaged word embeddings is an effective model for retrieving Pubmed
abstracts. However, they do not compare against standard IR baselines such as BM25, which makes
an interpretation of their finding difficult.

We argue that instead of relying on the cheap cosine similarity it may be better to employ an expensive but more accurate similarity function.
The exact brute force search using this function would be expensive,
but the cost could be reduced by applying an approximate search method for generic---i.e., not necessarily metric---spaces.

A common approach to non-metric space indexing involves
projecting data to a low-dimensional Euclidean space.
The goal is to find a mapping without a large distortion of the original similarity measure.
Jacobs et al.~\cite{JacobsEtAl2000} review projection methods
and argue that such a coercion is often against the nature of a similarity measure,
which can be, e.g., intrinsically non-symmetric.

%
% This is in fact no true, you can symmetrize efficiently.
% It's not clear though if the resulting method will be effective.
%

%In the case of non-symmetric IBM \modelonens, symmetrization may be a reasonable thing to do,
%but it can be prohibitively expensive.
%As explained in \S~\ref{sec:model1}, to speed up computation of $P(Q|A)$, we store translation probabilities
%associated with query terms in the form of the inverted index. 
%The resulting index is large and expensive to compute.
%Should we decide to symmetrize \modelone scores in a straightforward way, i.e., by computing $0.5\times(P(Q|A)+P(A|Q))$,
%we need to create an additional index for each answer, which is not practical.
%Yet, without the index, computation of $P(A|Q)$ becomes nearly an order of magnitude slower.

Among other factors, the lack of symmetry prevents us from
using the kernelized LSH \cite{kulis2009kernelized,Mu2010NonMetricLH}.
The only LSH variant that might be directly applicable in our case is the Distance-Based Hashing (DBH) \cite{athitsos2008nearest},
which uses randomly selected pivots to project points to a one-dimensional space via FastMap \cite{Faloutsos1995}. 
The space is further binarized so that approximately one half of data points are mapped to one,
and the other half is mapped to zero. 

While a detailed comparison of pivoting approaches to DBH is out the scope of the paper, 
we hypothesize that performance of DBH---like performance of NAPP---depends on the choice of pivots. 
In the case of NAPP, we have found that composing pivots from randomly selected terms allows us to achieve \emph{substantially} better
performance than selecting pivots randomly.
Thus, engineering pivots to support effective searching in a non-metric space seems to be an important research area.
Results obtained from this area will likely benefit both DBH and NAPP.

Proximity graphs (see \S~\ref{sec:knn-search}) is another promising class of 
distance-based methods, which are shown to be useful in non-metric spaces \cite{NaidanBN15}.
In this work we employ the SW-graph \cite{malkov2014approximate},
which works quite well for dense vector spaces. However, it has been less useful for 
\bmtf and \bmmodelonens. We have not been able to understand what causes the lack of performance,
but this remains an important research question as well.

Finally, we want to highlight the relationship of our approach to indexing automatically learned features for QA \cite{Yao2013}.
Yao et al.  propose to automatically learn associations between a question type
and various linguistic annotations such as named entities and POS tags \cite{Yao2013}.
For example, for a question  ``Who is the president of the United States'' an answer sentence
contains a person name (a named entity). Given a training corpus, we can automatically learn
associations and exploit them to guide the retrieval process. Technically, this requires
indexing linguistic annotations and carrying out a query expansion by adding expected annotations to the query.

For efficiency reasons, this works well only if we can find few strong associations for a query.
To demonstrate that this is not true in the case of the vocabulary gap, we compute effectiveness of \bmmodelone 
for varying sizes of the translation table.
Specifically, we prune all the entries $T(q|a)$ below a threshold. 
In addition, we estimate the average number of non-zero translation entries $T(q|a)$ associated with a \emph{single} query term.
As a reference point we also include data for \bmtfns.
We present results only for \comprns, because results for \stacko are analogous.

\begin{table}[h]
\small
\centering

\begin{tabular}{lll}
\toprule
\multicolumn{1}{c}{\begin{tabular}[c]{@{}c@{}}Minimum \\ translation \\ probability\end{tabular}} & \multicolumn{1}{c}{P@1} & \multicolumn{1}{c}{\begin{tabular}[c]{@{}c@{}}Number of \\ associated \\ terms\end{tabular}} \\ \midrule
BM25                                                                                              & 0.065                   & N/A                                                                                          \\
0.1                                                                                               & 0.066 (+2.6\%)          & 1700                                                                                         \\
0.05                                                                                              & 0.070 (+8.6\%)          & 3800                                                                                         \\
0.025                                                                                             & 0.073 (+12.5\%)         & 6200                                                                                         \\
0.005                                                                                             & 0.077 (+19.3\%)         & 12000                                                                                        \\
0.0025                                                                                            & 0.079 (+21.6\%)         & 15000                                                                                        \\ \bottomrule
\end{tabular}

\caption{Average number of terms associated with a query term at various performance levels of \bmmodelone (estimated on \emph{dev2}).
The first row represents a \bmtf run.
\label{tab:assoc_qty}
}
\end{table}

According to Table \ref{tab:assoc_qty},
outperforming \bmtf by about 20\% requires to keep more than 10K associations per query term (on average).
This number is so high because frequent words, which tend to appear in queries and text,
are associated with many less frequent words (i.e., respective translation probabilities are non-zero).

If we keep only translation entries with high ($\ge0.1$) probabilities,
the improvement over BM25 is merely 2.6\%. Yet, we still have to keep nearly 2K associations per query term!
This further corroborates the finding of Furnas~et~al.~\shortcite{furnas1987vocabulary} that 
accurate retrieval requires 
using a large number of term aliases, which is hard to implement using term-based indices. 
Yet, it is possible to do within a framework of the \knn search.

That said, the proposed methods are likely have limitations as well. For example, for both data sets employed in our 
experiments, the queries are quite long. It is not yet clear if \knn can be applied to shorter ad hoc queries,
which are frequently submitted to Web search engines.

%\leoinline{Proximity graphs:}
%\cite{toussaint1980relative,arya1993approximate,sebastian2002metric,houle2005fast,hajebi2011fast,dong2011efficient,wang2015fast}.

%\leoinline{For the SMT QA lit review}
%\cite{Berger2000,Echihabi2003,Soricut2006,RiezlerEtAl2007,Xue2008,surdeanu2011learning,Fried2015}.

%% file: conclusion.tex
In this paper we attempt to replace the classic term-based retrieval with the \knn search.
To this end, we train a linguistically motivated \emph{non-metric} and \emph{non-symmetric} similarity function:
a weighted combination of BM25 scores and IBM Model~1 log-scores.
Then, we demonstrate that it is possible to carry out an \emph{efficient} and \emph{effective} approximate \knn search using this function.

An exact brute-force \knn search using this similarity function is slow. Yet, an approximate algorithm can be nearly two orders of magnitude faster at the expense of only a small loss in accuracy. A retrieval pipeline using an approximate \knn search can be sometimes both faster and more  accurate compared to the term-based Lucene pipeline (see Table~\ref{tab:main}).
The success of our approach stems from the novel combination of existing methods and new algorithmic tricks to compute IBM Model~1 efficiently.

While the \knn search has been previously applied to IR and NLP problems \cite{petrovic2010streaming,Ture2011,Li2014,Petrovic2012,Moran2016,brok2016},
the previous work focuses largely on the cosine similarity and LSH methods (see \S~\ref{sec:discuss} for a discussion). 
This is the first successful attempt to apply a \emph{generic} \knn search algorithm
to a similarity function as challenging as a combination of BM25 and IBM Model~1.
In that, we find that the cosine similarity alone (in particular, the cosine similarity between averaged word embeddings) lacks a lot in effectiveness (see Figure~\ref{fig:recall_all}).

The focus of our study is on techniques that bridge the vocabulary gap.
Yet, our methods are generic in the sense that they can be used to model various types of semantic and syntactic mismatch~\cite{bilotti2007structured,Yao2013}. This opens up new possibilities for designing effective retrieval pipelines.
%While being generic our methods may have limitations. For example, it is not yet clear if they are applicable to short queries.

Our software (including data-generating code) and derivative data based on
the \stacko collection is available online.\footnote{\anonymizeurl{https://github.com/oaqa/knn4qa}}